\documentstyle[12pt]{article}

\begin{document}
{\sf \begin{center} \noindent {\Large \bf Effective black holes from non-Riemannian vortex acoustics in ABC flows}\\[3mm]

by \\[0.3cm]

{\sl L.C. Garcia de Andrade}\\

\vspace{0.5cm} Departamento de F\'{\i}sica
Te\'orica -- IF -- Universidade do Estado do Rio de Janeiro-UERJ\\[-3mm]
Rua S\~ao Francisco Xavier, 524\\[-3mm]
Cep 20550-003, Maracan\~a, Rio de Janeiro, RJ, Brasil\\[-3mm]
Electronic mail address: garcia@dft.if.uerj.br\\[-3mm]
\vspace{2cm} {\bf Abstract}
\end{center}
\paragraph*{}
It is shown that the existence of effective black holes in
non-Riemannian vortex acoustics is not forbidden under certain
constraints on acoustic torsion vector. In Arnold-Beltrami-Childress
(ABC) flows only the spatial part acoustic torsion vertical vanish,
which allows us to consider constraints on acoustic torsion to
obtain artificial non-Riemannian black hole solutions in ABC flows.
An explicity acoustic black hole solution is given for irrotational
ABC flows, This solution is similar to Visser acoustic metric for a
vortex flow. Actually a particular form of this acoustic metric is
given for ABC flows. Effective BHs with acoustic torsioned vortex
flows in effective spacetime. The Ricci scalar computed is the same
as that of a string singularity and the black-hole ergoregion
coincides with the event horizon.{\bf PACS
numbers:\hfill\parbox[t]{13.5cm}{02.40.Hw:differential geometries.
}}}

\newpage
\newpage
 \section{Introduction}
 Previously the author \cite{1} has built a non-Riemannian vortex acoustics theory of effective gravity \cite{2} which generalised the Unruh
 Visser \cite{3,4} effective theory of artificial black holes \cite{4} in other laboratory frameworks as optical and acoustic media as well as more recently
 plasma matter \cite{5}. These acoustic black holes are effective
 analogue pseudo-Riemannian metrics given by the homogeneous wave equation from linearised Euler flows or maximum Navier-Stokes. Non-Riemannian
 vortex acoustic metrics, of course, cannot be reduced to
 Unruh-Visser (UV)
 under no constraint acoustic BHs when Cartan torsion vector vanishes therefore
 expression ${(18)}$ in reference \cite{1} is not wrong and only the expression for the time component of acoustic torsion trace is wrong. In this paper besides
 of correcting this mistake we work it out the its consequences. First it is shown that even the non-vorticity flows sonic BHs may
 exist in non-Riemannian effctive spacetime. A non-diagonal sonic metric
 is obtained in this effective spacetime.
 Constraining the equations in non-Riemannian flows \cite{5}
 where the torsion fluctuation is orthogonal
 to the perturbed flow, it is shown that Riemannian sonic black holes can be also obtained. ABC flow and its effective spacetime are also considered. \newpage
\section{Acoustic Black-holes in effective
Riemann-Cartan ABC flows} In this section we present the brief
results described above. The non-Riemannian acoustic effective
geometry endowed with Cartan torsion , usually called Riemann-Cartan
effective spacetime. These equations which shall be used to build
the generalized acoustic BHs, are given by the force equation
\cite{1}
\begin{equation}
{\rho}[{\partial}_{t}{\textbf{v}}+({\textbf{v}}.{\nabla})\textbf{v}]=-{\nabla}p
\label{1}
\end{equation}
 $\textbf{v}$ is only partially irrotational
\begin{equation}
\textbf{v}_{1}={\nabla}{\psi}_{1} \label{2}
\end{equation}
which yields
\begin{equation}
\vec{\Omega}_{1}={\nabla}{\times}{\nabla}{\psi}_{1}=0 \label{3}
\end{equation}
where ${\Omega}_{1}$ is the vorticity fluctuation according to the
rules the fields fluctuations \cite{4}
\begin{equation}
p=p_{0}+{\epsilon}p_{1}\label{4}
\end{equation}
\begin{equation}
{\psi}={\psi}_{0}+{\epsilon}{\psi}_{1}\label{5}
\end{equation}
\begin{equation}
{\rho}={\rho}_{0}+{\epsilon}{\rho}_{1}\label{6}
\end{equation}
\begin{equation}
{\vec{\Omega}}={\vec{\Omega}}_{0}+{\epsilon}{\vec{\Omega}}_{1}\label{7}
\end{equation}
and the conservation mass equation
\begin{equation}
{\partial}_{t}{\rho}+{\nabla}.({\rho}{\textbf{v}})=0 \label{8}
\end{equation}
Along with the barotropic equation of state
\begin{equation}
p=p({\rho})\label{9}
\end{equation}
where p is the pressure, substitution of the above fluctuations into
the evolution flow equations one obtains the generalised Visser
equation as
\begin{equation}
{\partial}_{t}[{c^{-2}}_{sound}{\rho}_{0}({\partial}_{t}{\psi}_{1}+\textbf{v}_{0}.{\nabla}{\psi}_{1})]
+{\partial}_{t}[{c^{-2}}_{sound}{\rho}_{0}{\delta}+{\alpha}]]=
{\nabla}.[{\rho}_{0}{\nabla}{\psi}_{1}-{c^{-2}}_{sound}{\rho}_{0}\textbf{v}_{0}({\partial}_{t}{\psi}_{1}+\textbf{v}_{0}.{\nabla}{\psi}_{1})+{\alpha}]
\label{10}
\end{equation}
from the RC torsion one obtains the wave equation
\begin{equation}
{\Box}{\psi}_{k}=-{T^{l}}_{kl}{\psi}^{k}\label{11}
\end{equation}
where ${T^{l}}_{kl}$ is the torsion tensor trace. From the above
equations one obtains
\begin{equation} T^{0}=-\frac{{\rho}_{0}}{{c_{sound}}^{2}} \label{12}
\end{equation}
and \cite{1}
\begin{equation}
\textbf{T}=\frac{{\rho}_{0}}{{c_{sound}}^{2}}{\Omega}_{0}{\times}{\textbf{v}_{0}}
\label{13} \end{equation} actually formula (\ref{12}) corrects
previous results in reference \cite{1}. Actually to have the torsion
trace compute as above one has to impose on the unperturbed flow
before fluctuation, because the RHS of equation (\ref{10}) can be
expressed as
\begin{equation}
({\nabla}.{\textbf{v}_{0}}){\alpha}-({\textbf{v}_{0}}.{\nabla}){\alpha}
\label{14}
\end{equation}
and only the second in this expression contributes to acoustic
torsion. Therefore to be compatible with Unruh metric plus acoustic
torsion the first term in (\ref{14}) has to vanish and this
corresponds to the situation that either
\begin{equation}
{\nabla}.{\textbf{v}_{0}}=0 \label{15}
\end{equation}
which means incompressibility and the second condition namely that
${\alpha}$ vanishes means. Of course this condition shows that when
the unperturped fluid is an ABC flow, the sonic BHs are actually
Riemannian, but since this is impossible due the fact that ${T^{0}}$
cannot vanish, the sonic BH in ABC unperturbed flow has to be
incompressible in this effective spacetime. Now let us apply these
results in ABC flows. In these Beltrami like flow the vorticity is
proportional to the own velocity flow as described in the equation
\begin{equation}
\vec{\Omega}_{0}={\nabla}{\times}{\textbf{v}_{0}}={\lambda}{\textbf{v}_{0}}
\label{16}
\end{equation}
where ${\lambda}$ is constant. Thus when only the unperturbed part
of the flow is ABC's fluid one immeadiatly notes from expression
(\ref{13}) that acoustic torsion spatial part vanishes. However ,
the time component of torsion trace build as a vector, does not
vanish and effective spacetime is still non-Riemannian.  Now by
dropping the exigence of ABC flows and considering irrotational
incompressible flows one may write
\begin{equation}
{\nabla}.{\textbf{v}_{0}}={\nabla}^{2}{\psi}_{0}=0 \label{17}
\end{equation}
This can have a solution
\begin{equation}
{\psi}_{0}=\frac{A}{r} \label{18}
\end{equation}
and by the effective metric the Riemannian acoustic metric given by
\begin{equation}
g^{00}=-\frac{1}{{c^{2}}_{sound}{\rho}_{0}}\label{19}
\end{equation}
\begin{equation}
g^{0j}=-\frac{1}{{c^{2}}_{sound}{\rho}_{0}}{v^{j}}_{0}\label{20}
\end{equation}
\begin{equation}
g^{ij}=\frac{1}{{c^{2}}_{sound}{\rho}_{0}}({c}_{sound}({\delta}^{ij}-{v_{0}}^{i}{v_{0}}^{j}))
\label{21}
\end{equation}
one obtains a Riemannian black hole solution as
\begin{equation}
g^{0j}=\frac{1}{{c^{2}}_{sound}{\rho}_{0}}\frac{A}{r^{2}}{{\delta}^{j0}}\label{22}
\end{equation}
and
\begin{equation}
g^{ij}=\frac{1}{{c^{2}}_{sound}{\rho}_{0}}({c}_{sound}({\delta}^{ij}-\frac{A^{2}}{r^{4}}{\delta}^{ij}))
\label{23}
\end{equation}
The Riemannian line element is given by
\begin{equation}
d{s_{2+1}}^{2}=-({c_{sound}}^{2}-\frac{A^{2}}{r^{2}})dt^{2}+(dr-\frac{A}{r}dt)^{2}+r^{2}d{\theta}^{2}
\label{24}
\end{equation}
which possesses a curvature Ricci scalar like
\begin{equation}
R_{ABC}=-\frac{4{c_{sound}}^{2}{A^{2}}}{r^{4}} \label{25}
\end{equation}
which coincides with the string metric curvature scalar computed by
Visser \cite{6}. In our case the ergo-region coincides with the
event horizon since
\begin{equation}
r_{ergo-region}=\frac{A}{{c^{2}}_{sound}} \label{26}
\end{equation}
Therefore one finally obtains a new solution of sonic Riemannian
black holes. A non-Riemannian solution can be given by considering a
solution such as $\textbf{v}_{0}=constant$ which reduces the torsion
trace to a zero spatial part and the flow is irrotational. Thus one
concludes that it is possible to have a stationary unperturbed flow
without having a vanishing torsion trace, which means that
non-Riemannian sonic black holes exists even for non-vortex flow.
\newpage
\section{Conclusions}
 Recently a deep connection between vortex flows and non-Riemannian artificial acoustic BHs and the
by Garcia de Andrade \cite{1}. In this report one shows that this
sonic non-Riemannian BHs may exists even in the absence vorticity.
Other models in Einstein-Cartan gravity \cite{5} may be suitable for
finding other solutions of acoustic BHs.
\section{Acknowledgements} Special thanks go to Matt Visser which pointed out the mistake in reference \cite{1},
which motivate the previous comments and new proposals. I thank
financial supports from Universidade do Estado do Rio de Janeiro
(UERJ) and CNPq (Brazilian Ministry of Science and Technology).

  \end{document}